# EEG decoding with conditional identification information

**Pengfei Sun [1], Jorg De Winne[1], Paul Devos[1] and Dick Botteldooren[1]**

[1] Department of Information Technology, WAVES Research Group, Ghent University, Gent, Belgium
E-mail: {pengfei.sun, jorg.dewinne, p.devos, dick.botteldooren} @ugent.be

**Summary:** Decoding EEG signals is crucial for unraveling human brain and advancing brain-computer interfaces. Traditional machine learning algorithms have been hindered by the high noise levels and inherent inter-person variations in EEG signals. Recent advances in deep neural networks (DNNs) have shown promise, owing to their advanced nonlinear modeling capabilities. However, DNN still faces challenge in decoding EEG samples of unseen individuals. To address this, this paper introduces a novel approach by incorporating the conditional identification information of each individual into the neural network, thereby enhancing model representation through the synergistic interaction of EEG and personal traits. We test our model on the WithMe dataset and demonstrated that the inclusion of these identifiers substantially boosts accuracy for both subjects in the training set and unseen subjects. This enhancement suggests promising potential for improving for EEG interpretability and understanding of relevant identification features.

## 1. Introduction

The interplay between humans and artificial intelligence (AI) remains suboptimal, lacking the depth of engagement and synchrony inherent to human-to-human interactions. In pursuit of bridging this gap, there has been a marked shift towards leveraging neurophysiological insights, particularly through the prism of electroencephalography (EEG), to elucidate underlying cerebral mechanisms and refine the human-computer interface. The WithMe [1] experiment exemplifies this approach by presenting subjects with specific auditory and visual stimuli, thereby enabling the differentiation between target and distractor stimuli, whilst concurrently capturing the resultant EEG data. However, another challenge that arises is decoding the collected EEG signals, and in particular how to effectively decode and analyze the data to extract meaningful information.

Recently, advancements in machine learning have shown notable advantages in extracting intricate information from EEG signals [2][3]. Among these techniques, Convolutional Neural Networks (CNNs) and Recurrent Neural Networks (RNNs) stand out. CNNs process EEG signals as frames, synthesizing this data to make final decisions. RNNs, in contrast, retain information from previous inputs, showcasing an ability to recognize and remember temporal sequences, which is crucial for tasks needing long short-term memory of past events. Initial explorations employing deep neural networks (DNN) and conventional machine learning paradigms have yielded promising outcomes by directly processing EEG signals from WithMe experiment to classify the target/distractor [4]. The majority of neural network solutions for EEG decoding utilize fully supervised learning methods, meaning they refine their parameters based on hard-labeled data. However, this method tends to create models that are highly specialized for the tasks they're trained on, which may not perform well on different tasks or with new individuals [5]. In addition, the heterogeneity in individual brain activity patterns poses a significant challenge to the current deep learning frameworks, particularly in decoding EEG signals from subjects not represented in the training corpus.

Notwithstanding these challenges, the WithMe experiment has unveiled certain individual characteristics, notably sensory dominance [6], that substantially influence experimental outcomes. For instance, participants with auditory dominance exhibited superior performance across various metrics and conditions compared to their visually dominant peers [1]. This observation prompts a reevaluation of the role individual-specific traits play in modulating EEG signals in an attention and working memory task. It raises the intriguing possibility that integrating a compendium of these personal attributes into computational models could potentially enhance their representational capacity. By decoding the latent interplay between personal traits and EEG patterns, this research aspires to not only bolster decoding accuracy for familiar subjects but also extend predictive proficiency to novel individuals.

To tackle this issue and recognize that personal characteristics can impact experiment results, we propose a novel framework that incorporates conditional identification information into the EEG decoding process. Thus, a network employing fully supervised learning can utilize not only the hard label information but also the conditional identification information. This paper is dedicated to investigating the viability of this innovative methodology, with the ultimate aim of advancing human-AI interactions.

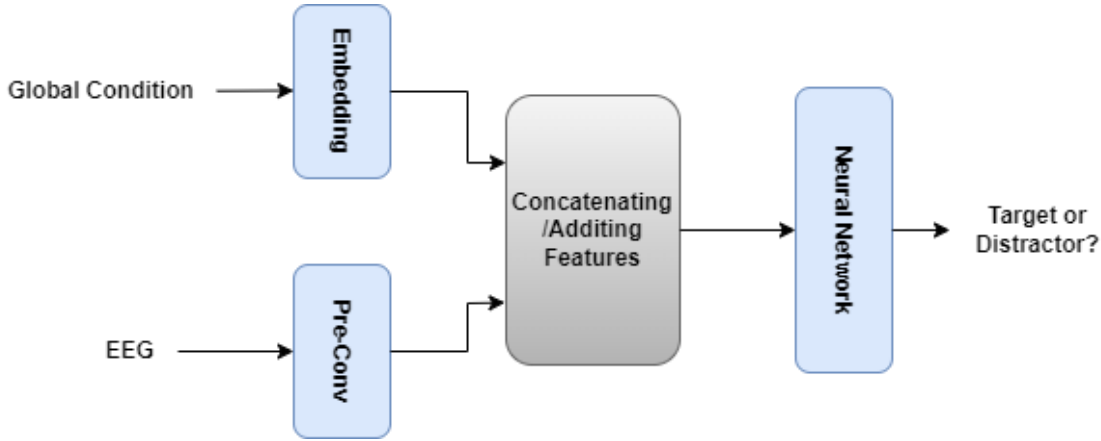

**Fig.1.** Overview of the Proposed Framework.

## 2. Overview

### 2.1. Overview of framework

Figure 1 depicts the structure of our proposed framework, comprising two main components: (1) Embedding the conditional identification information, employing a 16-neuron embedding layer designed to transform conditional identification information. Alongside, the pre-convolution layer, functioning as an identity layer in this study, encodes EEG data. (2) Decoding the integrated features, where this section is capable of utilizing various renowned neural network models to distinguish effectively between target and distractor stimuli.

### 2.1. Conditional identification information

The conditional identification information of each individual is utilized for target/distractor classification. This auxiliary conditioner employs an embedding layer to encode the identification attributes of each subject, transforming them into a comprehensive subject embedding. These embeddings, along with the EEG patterns, are then synergistically fused and introduced into the neural network. Through this extension, we expect to enhance the model's capability to learn a more generalized representation across diverse individuals, more precisely accounting for the variability in their brain activity characteristics. In this paper, we choose four distinct variables to examine their influence on the outcomes: 'Auditive/Visual Dominance', 'Sex', 'Music Education', and 'Active musician'. The former is assessed with the experiment proposed in [6], the others can be obtained via a simple questionnaire.

## 3. Experiment and Results

### 3.1. Dataset

Our model was trained and evaluated using the dataset from the WithMe experiment [1]. This experiment presented target and distractor digits to the subject asking them to remember and rename the targets. Four conditions were tested: simple sequence of visual stimuli, rhythmic presentation of targets, simple presentation supporting targets with a short beep, rhythmic presentation supported by beeps. The dataset encompasses data from a total of 42 participants. For training and internal testing, we randomly selected 38 participants further referred to as Within-subjects. The remaining 4 participants' data were reserved to assess the generalizability of the models and are referred to as Unseen-subjects. Specifically, we partitioned the WithMe data into a training set and two testing sets: Within-subjects, which comprises 18,176 training instances and 4,580 testing instances, and Unseen-subjects, which includes 2,400 testing instances. Preprocessing of the EEG data involved re-referencing each channel to the average activity of the mastoid electrodes. The data were then band-pass filtered between 1 and 30 Hz and subsequently downsampled to 64 Hz. Then, the data were segmented into 1.2 s epochs based on trigger events, with the final preprocessing step normalizing the EEG channel data to ensure zero mean and unit variance for each sample. The data and code can be accessed via https://github.com/sunpengfei1122/Withme-EEG-dataset.

**Table 1.** The results of three models on WithMe dataset is presented and compared to the models with the global condition id information.

| Datasets | Models | Within Accuracy | Unseen Accuracy |
|---|---|---|---|
| WithMe | EEGNet | 81.67% | 76.42% |
| | + IDs | **86.29%** | **79.08%** |
| | LSTM | 80.09% | 74.00% |
| | + IDs | **81.18%** | **76.00%** |
| | DMU | 81.94% | 75.92% |
| | + IDs | **82.21%** | **77.21%** |

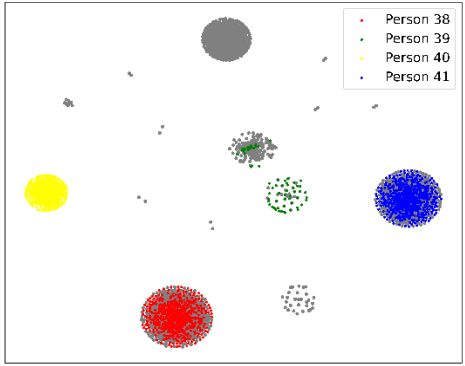

**Fig. 2.** Illustration of the intrinsic clustering pattern of identification information after embedding layer, as unveiled by t-SNE. The grey clustering pattern represents all the trained subjects, while the colorful pattern denotes the unseen subjects.

### 3.2. Implementation detail

In our experiment, we use the Adam optimizer to optimize the weights with a constant learning rate of 0.0001 and a minibatch size of 128. The EEGNet architecture features convolutional layers, starting with 16 kernels for initial temporal and spatial feature extraction from EEG signals. This is followed by depthwise and separable convolutions using 32 and 64 kernels, respectively, for efficient feature learning. For the LSTM [7] and DMU [8] models, a single recurrent unit with 64 neurons is utilized. Specifically, for the DMU's delay gate, the total number of delays is set to 20, considering the short duration of each sample. These models are developed within the PyTorch framework, adhering to default training methodologies. All modules are trained and updated in an end-to-end manner.

### 3.3. Results

Table 1 delineates the performance of three baseline models (EEGNet, LSTM, and DMU) and their counterparts incorporating our conditional identification (IDs) information branch of each participant. Remarkably, the EEGNet model, when enriched with conditional information, exhibits substantial enhancements in performance in both within-subject and unseen-subject. Furthermore, the addition of conditional IDs to LSTM and DMU models also yields marked improvements, particularly in the recognition of unseen subjects, indicating that the network has acquired more generalized representation of EEG. Additionally, t-distributed Stochastic Neighbor Embedding (t-SNE) [9] visualizations across all individuals of the conditional identification embedding layer in Figure 2 reveal a tendency for unseen subjects (person 38 to 41) to gravitate towards familiar centroids. Intriguingly, while up to 14 cluster centers (according to experimental data statistics) are theoretically possible given the 4-dimensional input IDs, only 7 prominent clusters emerge, suggesting that not all features exert a significant influence on the model's performance.

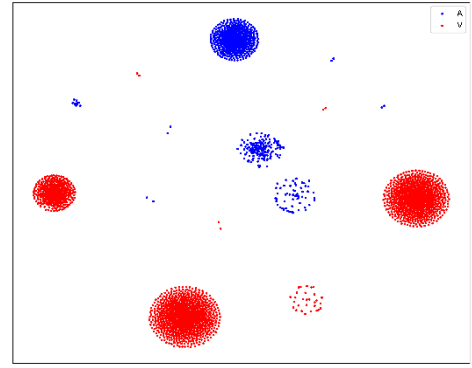

**Fig. 3.** Illustration of the intrinsic clustering pattern of Audio/Visual (A/V) dominance.

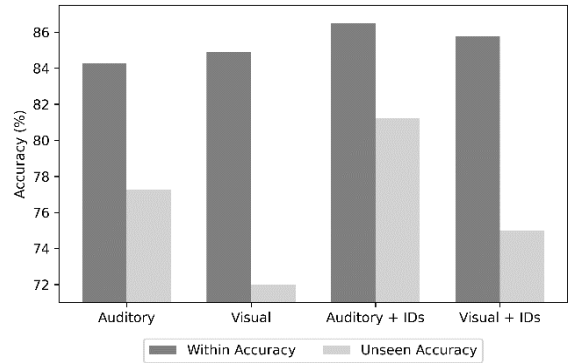

**Fig. 4.** Classification Performance on the WithMe Dataset based on Auditory/Visual dominance.

### 3.4. Analysis

To further validate our proposed framework, we focus on a key personal trait: Dominance. The WithMe study [1] demonstrated that participants with auditory dominance outperformed visually dominant individuals in all metrics and scenarios. As illustrated in Fig. 3, this distinction is shown as two distinct clusters based on dominance type. Subsequently, we evaluate EEGNet in two contexts: Auditory vs. Visual dominance for EEG classification. Fig. 4 reveals that, for the vanilla EEGNet, visually dominant individuals slightly outperform their auditory counterparts in within-subject tests but fare worse in unseen situations. However, when incorporating IDs information, the auditory group excels in both scenarios, consistent with our experimental findings. This observation may suggest that participants who performed better in the experiment tended to have clearer representations in their EEG signals that can be recognized by neural networks.

## 4. Conclusions

In this paper, we investigate the effectiveness of incorporating additional conditional identification information into neural network architectures for the classification of target versus distractor stimuli based on EEG. Through the deployment of an auxiliary global conditioner that utilizes an embedding layer to capture unique individual traits, our methodology not only enhances the model's precision in the same subjects but also amplifies its generalizability to unseen subjects, adeptly navigating the variety of neural responses observed in diverse individuals. Our results suggest that incorporating a personalized and context-aware conditioner is a promising approach to enhance the performance and reliability of EEG classification in real-world scenarios.

## Acknowledgements

This work was supported in part by the Flemish Government under the "Onderzoeksprogramma Artificiele Intelligentie (AI) Vlaanderen" and the Research Foundation - Flanders under grant number G0A0220N (FWO WithMe project).